\documentclass[runningheads]{cl2emult}

\usepackage{makeidx}  
\usepackage{epsfig}
\usepackage{graphicx} 
\usepackage{subeqnar} 
\usepackage{multicol} 
\usepackage{cropmark} 
\usepackage{eso}      
\makeindex            

\usepackage{amssymb}

\newcommand{\decdeg}[2]{$#1\mbox{$^\circ\mskip-6.6mu.\,$}#2$}

\begin{document}
\title*{Finding the First Stars: The Hamburg/ESO Objective Prism Survey}

\toctitle{Finding the First Stars:
\protect\newline The Hamburg/ESO Objective Prism Survey}
%
%
\titlerunning{Finding the First Stars}
%
\author{
  Norbert Christlieb\inst{1}
  \and Dieter Reimers\inst{1}
  \and Lutz Wisotzki\inst{1}
  \and Johannes Reetz\inst{2}
  \and Thomas Gehren\inst{2}
  \and Timothy C. Beers\inst{3}
  }
\authorrunning{Norbert Christlieb et al.}
%
%
\institute{
  Hamburger Sternwarte, Germany
  \and Universit\"ats-Sternwarte M\"unchen, Germany
  \and Department of Physics \& Astronomy, Michigan
  State University, U.S.A.
  }

\maketitle              

\section{The Hamburg/ESO survey}

The Hamburg/ESO survey (HES; [4]) is an objective-prism survey for bright
quasars based on IIIa-J plates taken with the ESO Schmidt telescope and its
4$^{\circ}$ prism. It covers the total southern extragalactic sky
($\delta<$\decdeg{+2}{5}; $|b|\gtrsim 30^\circ$). All 380 Schmidt plates
(effective area $\sim 7\,000$ square degrees) have been taken, and have been
digitized and reduced in Hamburg. The spectral range of the HES plates is
$3200\,\mbox{\AA} < \lambda < 5300\,\mbox{\AA}$, with a seeing-limited
spectral resolution of 15\,{\AA} at H$\gamma$. This makes it possible to
efficiently exploit the stellar content of the survey.

Among other interesting objects (e.g. white dwarfs, carbon stars, and field
horizontal branch A-type stars; [2]), extremely metal-poor stars can be
selected by automated procedures in the digital database of the HES. With its
magnitude range of $12 \gtrsim B_J \gtrsim 17.5$, the HES is more than one
magnitude deeper than the HK survey of Beers et al. [1]. Moreover, it covers
regions in the sky that are not covered by the HK survey. In result, the HES
can increase the total survey volume for metal-poor stars by more than a
factor of $4$ compared to the HK survey alone.

\section{Selection of metal-poor Stars}

We select metal-poor stars in the HES by means of automatic spectral
classification [3]. As a training set, we use an extensive grid of model
atmospheres converted to objective prism spectra. The grid covers the
following parameter range: $5200\,\mbox{K}<T_{\mbox{\footnotesize
    eff}}<6800\,\mbox{K}$ ($200$\,K steps), $2.2<\log g<4.6$ ($0.8$\, dex
steps) and $-0.3>\mbox{[Fe/H]}>-3.6$ ($-0.3$\,dex steps). Any object
brightness can be simulated by adding the appropriate amount of gaussian
noise. For automatic classification we apply a Bayes rule which employs
feature vectors $\vec{x}$ containing a large set of automatically detected
spectral features, e.g.  equivalent widths of stellar absorption lines, line
indices, broad band colours, and narrow band colours equivalent to Str\"omgren
$uvb\beta$.  Class-conditional probabilitiy distributions
$p(\vec{x}|\Omega_i)$ are modelled by multivariate normal distributions.

In a simulation we investigated the effective yield (EY) of our selection, in
dependence of $B$ magnitude. Since we currently concentrate our efforts on
metal-poor {\em turnoff\/} stars, we slightly modify the definition of EY
given by Beers (this volume):
\begin{equation}
{\rm EY}(y) = \frac{\mbox{\# stars with [Fe/H]}\le y\;\mbox{and }\log g\ge
  4.0}{\mbox{total \# of selected stars}}
\end{equation}
We used $\sim 30\,000$ spectra in each magnitude bin, distributed in [Fe/H]
according to a metallicity distribution function derived from the catalogue of
Beers et al. [1].

\begin{figure}
  \centering
  \epsfig{file=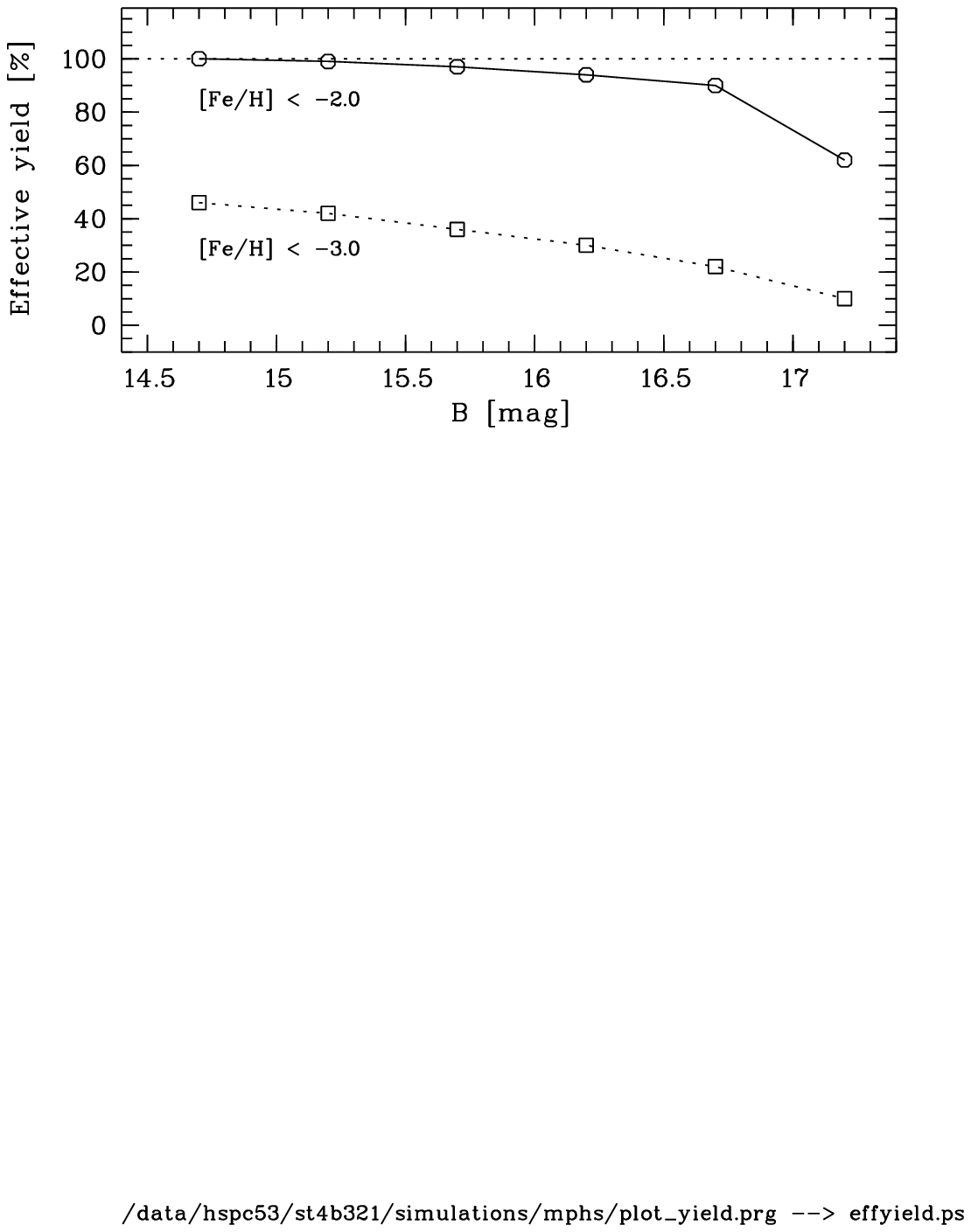, width=9cm, clip=, bbllx=72, bblly=510,
    bburx=371, bbury=656}
  \caption{\label{effyield} Effective yield of metal-poor turnoff stars
    selected by automatic classification in the HES, determined by simulations.}
\end{figure}

The result is shown in Fig. \ref{effyield}. Note that sample completeness and
EY are complementary. If the selection is optimized towards high EY, the
completeness is $\sim 60\,\%$ for $B<16.5$\,mag. If a completeness of $\sim
90\,\%$ is desired, EYs are typically by a factor $2$ lower.

Our results will be verified by spectroscopic follow-up observations in an
upcoming observing run at the ESO-NTT.

\clearpage
\addcontentsline{toc}{section}{Index}
\flushbottom
\printindex

\end{document}